\documentclass[12pt,preprint]{aastex}

\usepackage{amssymb,amsmath}
\usepackage{graphicx}
\shorttitle{Collisionally excited filaments in HH~1/2}
\shortauthors{Raga et al.}

\begin{document}

\title{The time evolution of HH~1 from four epochs of HST images}

\author{A. C. Raga\altaffilmark{1}, B. Reipurth\altaffilmark{2},
A. Esquivel\altaffilmark{1}, J. Bally\altaffilmark{3}}
\altaffiltext{1}{Instituto de Ciencias Nucleares, Universidad
Nacional Aut\'onoma de M\'exico, Ap. 70-543, 04510 D.F., M\'exico}
\altaffiltext{2}{Institute for Astronomy, University of Hawaii at Manoa, Hilo, HI 96720, USA}
\altaffiltext{3}{Center for Astrophysics and Space Astronomy, University of Colorado, UCB 389,
Boulder, CO 80309, USA}

\email{raga@nucleares.unam.mx}

\begin{abstract}
  We present an analysis of four epochs of H$\alpha$ and [S~II]~$\lambda\lambda$~6716/6731
  HST images of HH~1. For determining proper motions
  we explore a new method based on analysis of spatially degraded images obtained convolving the images
  with wavelet functions of chosen widths. With this procedure we are able to generate maps of proper motion
  velocities along and across the outflow axis, as well as (angularly integrated) proper motion velocity distributions.
  From the four available epochs, we find the time evolution of the velocities, intensities and spatial distribution
  of the line emission. We find that over the last two decades HH~1 shows a clear acceleration. Also, the
  H$\alpha$ and [S~II] intensities have first dropped, and
  then recovered in the more recent (2014) images. Finally, we show a comparison between the two available
  HST epochs of [O~III]~$\lambda$~5007 (1994 and 2014), in which we see a clear drop
  in the value of the [O~III]/H$\alpha$ ratio.
\end{abstract}


\section{Introduction}

The presence of high proper motions is one of the principal characteristics of Herbig-Haro (HH) objects,
and were observed in some HH objects even before they were identified as such (for example, in HH~29,
see Luyten 1963 and Cudworth \& Herbig 1979). The importance of proper motion measurements was
highlighted by the study of a time series of plates of HH~1 and 2 by Herbig \& Jones (1981), showing
that the two objects form part of a single, bipolar outflow (see the
discussion of the history of HH~1/2 proper motions of Raga et al. 2011).

An important issue about the proper motions of HH objects is the question of whether or not
they represent actual mass motions. George Herbig initially thought that these motions might
represent changes in the relative brightnesses of regions within concentrations of HH knots, but
in seeing the large displacements that were measured, he rapidly became
convinced that HH objects indeed had clear, substantial motions in the plane of the sky
(Kyle Cudworth, personal communication). Jones \& Herbig put together from their
plates (described in their 1981 paper) an ``animation'' of the motions of HH~1 and 2, which
convincingly show systematic, organized motions away from a central source.

The proper motions of HH objects are closely related to the variability of their emission.
Few studies of the variability of the emission line spectra of HH objects have been
made, because while the proper motions can be measured over time-periods of a few years,
the variability of the emission typically takes place over decades. The optical time-variability
of HH~1 and 2 has been studied by Herbig (1968, 1973) and Eisl\"offel et al. (1994), and
the UV time-variability has been described by Brugel et al. (1985). One of the few studies
of variability in other HH objects is the one of HH~29 by Liseau et al. (1996).

While the older studies of proper motions of HH objects utilized photographic
plates, more recent measurements use CCD images
(the first effort in this direction being the study of HH 1 and 2 by Raga et al. 1990a). The proper
motions of HH knots from CCD images
can be obtained from fits to intensity peaks (which requires an identification
of pairs of peaks in different epochs, see, e.g., Eisl\"offel et al. 1994).

It is also possible
to define ``boxes'' or ``apertures'' (i.e., angularly limited fields within the images) used
to calculate cross-correlation functions between pairs of images. The offsets associated with
the peak of the cross-correlation function give the value of the proper motion of the emission
within the chosen boxes. This method was first used in the context
of HH objects by Heathcote \& Reipurth (1992).
The cross-correlation method has the advantage of giving proper motions associated to possibly
well defined, larger features in HH objects, and is less sensitive to changes in the detailed,
small scale morphology of the flows.

In this paper, we use a new method for determining proper motions in HH objects:
\begin{itemize}
\item We first convolve the images of the successive epochs with a wavelet of half-width $\sigma$
  (with $\sigma\geq 1$~pixel),
\item In successive pairs of convolved images we identify neighbouring intensity peaks, and
  determine the proper motions from the resulting offsets.
\end{itemize}
This method has features in common with both the ``peak fitting'' method (indeed, it is a
peak fitting method, but using angularly degraded images) and the ``cross correlation''
method (as it includes spatial smoothing). The spatial smoothing feature is
convenient for proper motion measurements of extended structures (see the discussion above),
and the smoothing is obtained without having to manually choose arbitrary ``cross correlation
boxes''.

We carry out a first application of this method to 4 epochs of H$\alpha$ and [S~II]~$\lambda\lambda$~6716/6731
HST images of the HH~1/2 system, in particular focussing on the time-evolution of HH~1.
This data set is described in section 2. In section 3, we present the proper motions
derived from cross correlations of the brighter regions of HH~1. In section 4, we present
the proper motions of individual condensations of HH~1 obtained with the new ``wavelet
proper motion'' method. In section 5, we derive ``proper motion velocity distributions'' (as
a function of the velocities along and perpendicular to the outflow axis) from the
``wavelet proper motions'', and also the variability of the line intensities and the morphology
of HH~1 as a whole. Section 6 presents a qualitative comparison of the two available [O~III]~$\lambda$~5007
HST images of HH~1. Finally, the results are discussed in section 7.

\section{The data set}

Four epochs of images of the HH~1/2 outflow are now available in the HST archive (see Table 1).
The successive observations have been described in a series of papers:
\begin{itemize}
\item epoch 1: Hester et al. (1998),
\item epoch 2: Bally et al. (2002),
\item epoch 3: Hartigan et al. (2011),
\item epoch 4: Raga et al. (2015a, b).
\end{itemize}
The first three epochs were obtained with the WFCP2 camera, and the fourth epoch with the WFC3 camera.
The paper of Bally et al. (2002) presents proper motions obtained with the two first epochs, and
the paper of Hartigan et al. (2011) discusses the morphological changes seen in the
first three epochs. Using the  only two stars present in the frames, namely
the Cohen-Schwartz star (Cohen \& Schwartz 1979) and ``star no. 4'' of Strom et al. (1985)
we have centered, rotated and scaled all of the images, producing a set of aligned frames with
$0.1$~arcsec per pixel.

We obtain the fluxes of the individual pixels of the archival WFPC2 images by multiplying the ``data numbers''
(DNs) by a calibration constant $C$=BANDWID$\times$PHOTFLAM (with the values given by the corresponding
keywords in the fits files). We obtain $C=1.14\times 10^{-15}$, $4.51\times 10^{-16}$ and
$3.99\times 10^{-16}$~erg~s$^{-1}$~cm$^{-2}$, for the F502N, F656N and F673N filters, respectively
(see Table~1). Dudziak \& Walsh (1997) obtain calibration constants $C_{DW}=1.15\times 10^{-14}$
and $4.08\times 10^{-15}$~erg~s$^{-1}$~cm$^{-2}$, for the F502N and F656N filters
(they did not use the F673N) filter), which differ from our values by a factor
of almost one order of magnitude. We assume that this is due to a typo in the paper of
Dudziak \& Walsh (1997).

We obtain the fluxes of the WFC3 images using a calibration constant
$C=\Delta \lambda\times$PHOTFLAM (with the PHOTFLAM values given in the header
of the fits files, and the $\Delta\lambda$ ``rectangular width'' values given in the WFC3 Instrument
Handbook). We obtain $C=3.42\times 10^{-16}$, $2.94\times 10^{-16}$ and
$2.64\times 10^{-16}$~erg~s$^{-1}$~cm$^{-2}$, for the F502N, F656N and F673N filters, respectively
(see Table~1). Once allowance has been made for the angular size of the WFC3 pixels, these
calibration constants are found to be in good agreement with the values derived by
O'Dell et al. (2013). We finally bin the WFC3 images to the 0.1~arcsec pixel size
of the WFPC2.

In the F656N (H$\alpha$) filter, the [N~II]~$\lambda$~6548 line is present at $\sim 5$\%\  of the peak
transmission (see the discussion of O'Dell et al. 2013). Given the relatively low
[N~II] 6548/H$\alpha$ ratios observed in HH~1 and 2 (see, e.g., Brugel et al. 1981a),
the contribution of the [N~II] emission to the F656N frames is likely to be
only $\sim 2$\%\ . The continuum emission of HH~1 and 2 (Brugel et al. 1981b) probably
also has only a small contribution to the flux of the HST images (see the discussion
of Raga et al. 2015b).

\section{The H$\alpha$ and [S~II] proper motions of HH~1}

We use the H$\alpha$ and [S~II] images of the four epochs (see Figure 1) to obtain the proper motions
of the emitting region of HH~1. We first carry out cross correlations between the three possible
pairs of consecutive images (of each emission line) in order to determine the time-dependence of the
proper motions of HH~1.

We first carry out cross correlations over the field shown in the bottom right frame of Figure 1.
This field includes the brighter region of the head of HH~1.
From the three successive pairs of images, we determine
three proper motions for H$\alpha$ and for [S~II]. We then project the measured proper motions
along and across the outflow axis of the HH~1/2 system (for which adopt the PA=325$^\circ$ orientation
of the HH1 jet, see Bally et al. 2002). Assuming a distance of 414~pc, we then calculate the velocities
parallel ($v_\parallel$) and perpendicular ($v_\perp$) to the outflow axis for the three pairs of
images. The results of this exercise are shown in Figure 2. Actually, very similar results are
obtained if one carries out cross correlations over the whole field shown in Figure 1 (which is a
direct result of the fact that the head of HH~1 has a dominant contribution to the emission
of the object).

In Figure 2, we show the proper motion velocities along ($v_\parallel$) and across ($v_\perp$)
the outflow axis (with negative values of $v_\perp$ directed to the SW), as well
as the modulus $v_T=\sqrt{v_\parallel^2+v_\perp^2}$. We can see the following features:
\begin{itemize}
\item it is clear that $v_\perp$ has relatively small values, so that we always have
  $v_T\approx v_\parallel$,
\item $v_\parallel$ (and also $v_T$) shows a monotonic growth as a function of time,
  with an increase from $\approx 250$ to 300~km~s$^{-1}$ during the $\approx 20$~yr
  time-span of the observations,
\item $v_\perp$ has moduli $<100$~km~s$^{-1}$, and has alternating negative and positive
  values (though with considerably larger excursions in the SW, negative direction).
\end{itemize}

\section{The proper motions of individual condensations}

It is not straightforward to measure proper motions of individual features in
images of HH objects spanning many years. This is particularly true of high resolution
HST images, in which identifiable features have clearly visible morphological changes
(see, e.g., the evolution of the head of HH~1 in the frames shown in Figure 1). Because
of this, it has been standard practice to define arbitrary ``cross-correlation boxes'' (i.e.,
limited fields over which cross correlations between pairs of images are carried out), from
which proper motions are determined (corresponding to the shift of the cross-correlation
function). This method has the attractive feature that the implicit spatial smoothing (from carrying
out a cross-correlation over a relatively large-sized field) gives a proper motion which
is not sensitive to the small spatial scale morphological time-variations. However,
it has the undesirable feature that the measured proper motions are
dependent on the somewhat arbitrarily chosen cross-correlation boxes.

Alternatively, it is possible to evenly divide images into boxes (or ``tiles'', see
Szyszka et al. 2011) of a fixed size, and to carry out cross correlations within these boxes
(in pairs of successive images). This was first tried for HH objects by Raga et al. (2012).

In this paper we try a new method for determining proper motions:
\begin{itemize}
\item we first degrade the resolution of the original images (through a convolution with
  a smoothing function),
\item we then measure the shifts in the identifiable intensity peaks between pairs of
  spatially smoothed images. These shifts are measured through direct, paraboloidal fits to the
  intensity peaks in the images.
\end{itemize}
This method shares the spatial smoothing feature of the cross-correlation method,
and gives results which are not dependent on arbitrary choices of ``boxes''. However,
the obtained proper motions are indeed dependent on the width ($\sigma$, see below)
of the smoothing function.

For the smoothing function we choose a ``Mexican hat'' wavelet of
half-width $\sigma$:
\begin{equation}
  g_\sigma(x,y)=\frac{1}{\pi\sigma^2}\left(1-\frac{x^2+y^2}{\sigma^2}\right)\,
  e^{-(x^2+y^2)/\sigma^2}\,,
  \label{g}
\end{equation}
with which we convolve the observed frames ($x$ and $y$ are the coordinates in pixels
on the plane of the image). The convolved images $I_\sigma$ are then calculated through the usual integral
\begin{equation}
  I_\sigma(x,y)=\int\int I(x',y')\,g_\sigma(x-x',y-y')\,dx'dy'\,,
  \label{is}
\end{equation}
where $I(x',y')$ is the original (i.e., not convolved) image, and the convolution integral is carried
out over the whole domain with a standard, ``Fast Fourier Transform'' method.

We have chosen
two wavelets, with $\sigma=3$ and 5 pixels ($0.3$ and $0.5$~arcsec, respectively), and the
results of the convolutions of the H$\alpha$ images with these functions are shown in Figures 3 and 4.

The H$\alpha$ images convolved with the $\sigma=3$ wavelet (see Figure 3) show
a number of peaks. We have made a search for peaks which satisfy the following
criteria:
\begin{itemize}
\item intensities with values greater than $I_p=1.5\times 10^{-15}$~erg~s$^{-1}$~cm$^{-2}$~arcsec$^{-2}$. This
  value was chosen so that the main, identifiable knots are included, while rejecting
  fainter peaks that appear to be associated with noise,
\item centers with an individual pixel which has larger fluxes than all of the neighbouring
  pixels (including diagonal neighbours).
\end{itemize}
In this way we eliminate ``ridges'' as well as low intensity maxima. The peaks that are
identified (and fitted) in the convolutions of the four H$\alpha$ frames with the $\sigma=3$ wavelet
are shown in Figure 3. Errors in the positions of the peaks (resulting from the fitting procedure and
the centering of the images) are of the order of $\sim 0.2$~pix.

We search for peaks in the pairs of consecutive images as follows.
For the three possible pairs of consecutive images (1994-1997, 1997-2007
and 2007-2014, see Table~2 and Figures 1 and 3) we take the position of the peaks in the
earlier image, and search (in the NW quadrant only) for corresponding peaks in the latter
image of the pair. This search is made only to a maximum distance $d_{max}$, which
we have set to 10 pixels for the 1994-1997 image pair, to 30 pixels
for 1997-2007 and 20 pixels for 2007-2014 (corresponding to a maximum proper motion
velocity of $\approx 600$~km~s$^{-3}$ for the three image pairs). It is then necessary to check
``manually'' that the knot pairs chosen by the algorithm actually correspond to knot pairs that
appear to be the same physical feature. In this way, for each pair of images we obtain shifts
for the identified knot pairs.

With the above knot detection algorithm, in the four consecutive epochs (of H$\alpha$ images
convolved with a $\sigma=3$~pix wavelet) we detect 17, 14, 14 and 17 knots, respectively. We
choose 11, 5 and 14 knot pairs in the successive three pairs of epochs. The lower number of
pairs chosen for the second ($1997.58\to 2007.63$) pair of epochs is due to the larger morphological
changes that occur in this longer time-interval.

The bottom frame of Figure 3 shows the shifts of identified knot pairs in the 1994-1997
images (red arrows), the 1997-2007 images (blue arrows) and in the
2007-2014 images (purple arrows). We also show one arrow (in cyan) that corresponds
to the shift between leading NW intensity peak of the 2014.63 frame and the closest peak
to the SE found in the 2007 frame. This shift therefore corresponds to a ``backwards in time''
search for a companion knot in the 2007-2014 frame pair.

In Figure 4, we show the convolutions of the H$\alpha$ frames with a broader, $\sigma=5$ pixel ($0''.5$)
wavelet. These convolutions clearly show a smaller number of intensity peaks, with a less chaotic
distribution in the faint, SW region of HH~1. In these maps we find all of the peaks with central
intensities larger than $I_p=3.0\times 10^{-16}$~erg~s$^{-1}$~cm$^{-2}$~arcsec$^{-2}$, and identify
common knots in the pairs of successive frames (see the discussion above). The offsets between
the common knots in the three successive pairs of images are shown in the bottom frame of
Figure 4. In the bottom frame of this figure, the black arrow of the westernmost knot corresponds
to the offset between the 1997.58 and 2014.63 frames, because this region lies outside of
the ``HST footprint'' in the 2007.63 frames.

We now take the knot offsets measured in the images convolved with the $\sigma=3$ wavelet
(shown in the bottom frame of Figure 3) together with the offsets measured in the $\sigma=5$
convolutions (see Figure 4) in the S and SW region of HH~1 (which are not seen well in the
higher resolution maps) to produce proper motion velocities for the 1994-1997, 1997-2007
and 2007-2014 pairs of epochs. The results are shown in Figure 5 as maps of the velocities parallel
and perpendicular to the outflow axis.

These maps are generated as follows. For each pixel in the map we make a search for all of the knots with
measured proper motions within a radius $r_{max}$ (for each of the knot pairs, we use the mean knot position
of the two consecutive epochs). We then compute the velocity corresponding to the pixel
of the map as an average of the velocities of the neighbouring knots, with weights $w=r_{max}-r$, where
$r$ is the distance from the knot to the pixel under consideration.

The H$\alpha$ proper motions along the outflow axis for the three pairs of epochs (left
column of Figure 5) all show larger velocities along a central
``high velocity channel'' (as previously reported by Bally et al. 2002 from an analysis of the 1994 and 1997
frames), surrounded with lower velocity regions to the W and E. Further W, there is a higher velocity region
(corresponding to the faint filament seen in the W region of the H$\alpha$ frames, see Figure 1). It is clear that
there is a general trend of growing axial velocities as a function of time, with a rather dramatic
velocity increase in 2007-2014 at the leading, NW edge of HH~1.

The velocities across the outflow axis (right hand column of Figure 5) show a general expansion
away from the axis (with more positive velocities to the NE, and more negative values to the W). Also,
in the 1997-2007 proper motions we see more eastwardly directed velocities
than in the other two pairs of epochs.
Therefore, the results obtained from the proper motions of individual condensations of HH~1
are broadly consistent with the proper motions of HH~1 as a whole (see the previous section and Figure 2).

We convolved the four [S~II] images with a $\sigma=3$ wavelet (see above), and
identified knot positions and knot pairs in successive frames (with the same criteria that we used
for the $\sigma=3$ H$\alpha$ convolutions, see above). The [S~II] knot positions and offsets are
shown in Figure 6, and the resulting velocity maps (of velocities parallel and perpendicular to the outflow
axis) are shown in Figure 7. We have not used the results obtained from $\sigma=5$ pix wavelet
convolutions of the [S~II] images (as we have done for the case of the H$\alpha$ maps, see Figure
4) because they show basically the same structures that are seen in the higher resolution, $\sigma=3$ convolutions.

The high axial velocity, central channel is less clear in the [S~II] (Figure 7) than
in the H$\alpha$ (Figure 5) axial velocity maps. Also, the high velocity structure at the tip of HH~1
(see the 2007-2011 H$\alpha$ axial velocity map in Figure 5) is not seen in [S~II]. This is a direct result
of the fact that the protrusion seen in the 2014 H$\alpha$ frame (bottom left frame of Figure 1)
has very weak [S~II] emission, and does not produce a separate peak in the convolution of the [S~II]
frame with the $\sigma=3$ wavelet.

Except for these features, the H$\alpha$ (Figure 5) and [S~II] (Figure 7) velocity maps are similar.
The axial velocity maps show a trend of general acceleration. The velocities
perpendicular to the axis show more negative values (directed to the SW) in the 1994-1997 and 2007-2011 maps,
and more positive values in 1997-2007 for both emission lines. Finally, in the H$\alpha$ maps we seer
higher axial velocities in the faint SE region in 2007-2014 (see the bottom left frame of Figure 5).

\section{Line intensities and proper motion velocity distributions}

We have used the determinations of proper motions of individual condensations (seen in maps convolved
with Mexican Hat wavelets, see above) to obtain spatially integrated velocity distributions. To
do this, we have computed the fluxes in velocity bins of widths $\Delta v=60$~km~s$^{-1}$ by
adding the fluxes of the knots with proper motion velocities (either along or across the
outflow axis) falling within each bin. The fluxes of the individual knots (seen in the convolutions with wavelets)
are calculated as $F_{knot}=\pi\sigma^2F_{peak}$, where $F_{peak}$ is the peak intensity of
the knot, and $\sigma$ is the half-width of the wavelet (see equation 1).

In this way we obtain H$\alpha$ and [S~II] distributions as a function of either the
velocities parallel ($v_\parallel$) or perpendicular ($v_\perp$) to the outflow
axis. The distributions resulting from the three pairs of consecutive epochs are shown in Figure 8.

The axial velocity H$\alpha$ and [S~II] velocity distributions ($F$ vs. $v_\parallel$, left column of Figure 8)
show a dominant, high velocity intensity peak in the 1994-1997 and 1997-2007 epoch pairs at
$v_\parallel\approx 300$~km~s$^{-1}$. The 2007-2014 axial velocity distribution (bottom left frame
of Figure 8) is broader, extending to velocities of $\sim 500$~km~s$^{-1}$.

The H$\alpha$ and [S~II] distributions of flux versus velocity across the outflow axis $v_\perp$ (right column of
Figure 8) resemble each other in the 1994-1997 and 2007-2014 epoch pairs, but show substantial differences for
1997-2007. The 1997-2007 distributions show more flux at positive values of $v_\perp$ (i.e., directed to the NE)
than the other two pairs of epochs.

Table 2 gives the maximum and minimum values of $v_\parallel$ and $v_\perp$ obtained (in both the H$\alpha$
or the [S~II] knots) for each of the three pairs of epochs. From these values we calculate the
velocity ranges $\Delta v_\parallel$ and $\Delta v_\perp$ for the three epoch pairs (see Table 2).

If we assume that the condensations of HH~1 take part in a single, broken-up bow shock flow, these
results can be interpreted as follows.
Raga et al. (1997) showed that the total ranges of possible proper motion velocities along ($\Delta v_\parallel$)
and across ($\Delta v_\perp$) the outflow axis in a bow shock flow are both equal to the velocity $v_{bs}$ of the
bow shock relative to the pre-bow shock environment. However, given that in a bow shock flow one only
has a finite number of clumps, the observed ranges in proper motion velocities do not necessarily
sample the full possible velocity range. We therefore expect to have $\Delta v_\parallel,\,\,\Delta v_\perp
\leq v_{bs}$. From Table 2, we then see that the velocity ranges (along and across the outflow
axis) of the 1994-1997 and 1997-2007 pairs of epochs imply a lower boundary for the
bow shock velocity $v_{bs}\sim 350$~km~s$^{-1}$,
while the 2007-2014 epoch pair implies a higher, $v_{bs}\approx 500$~km~s$^{-1}$ value.

We have also measured the H$\alpha$ and [S~II] fluxes integrated over all of the emitting area of HH~1 (over
the $18\time 18$~arcsec field displayed in Figure 1). We have carried out a subtraction of the background
(even though this has
only a minor effect on the obtained fluxes), and obtained the H$\alpha$ and [S~II] fluxes for the
four epochs (see Table 2), which we show as a function of time in Figure 9.

The [S~II] flux shows a decrease as a function of time until 2007, with a
rate of $\approx 6$\%\ per year in the 1994-1997 and 2007-2014 periods, and
a smaller rate of less than 1\%\ per year during 1997-2007. This intensity drop appears
to be a continuation of the decrease of $\approx 4$\%\ per year in the [S~II] emission of the HH~1F
condensation during the 1987-1994 period measured by Eisl\"offel et al. (1994). From
2007 to 2014, the [S~II] flux {\it grows} by $\approx $\%\ per year (see Figure 9 and table 2)

The H$\alpha$ flux (see Figure 9) has drops of $\approx 8$\% per year during 1994-1997
and of $\approx 2$\% per year during 1997-2007. Then, during the 2007-2014
period, the H$\alpha$ flux has {\it grown} at a rate of $\approx 7$\%\  per year. The combination of
the H$\alpha$ and [S~II] variabilities leads to a [S~II]/H$\alpha$ ratio
that decreases from 1994 to 1997, and then increases monotonically from 0.7 to $\approx 1$
from 1997 to 2014.

Finally, we note that the angular extension of the HH~1 emission (along the outflow axis) has
a slight decrease from 1994 to 1997 and then monotonically grows
as a function of time (as can be seen in a qualitative way in the consecutive images shown in
Figure 1). In order to quantify this effect, we have calculated the angular length along the
outflow axis of the isophote corresponding to an intensity of $3\times 10^{-15}$~erg~cm$^{-2}$s$^{-1}$~arcsec$^{-2}$
in the H$\alpha$ and [S~II] frames. These lengths (calculated from the [S~II] and H$\alpha$ maps) are
shown as a function of time in Figure 10. It is clear that the angular size of HH~1 has grown
(by $\sim 20$\%\  in [S~II] and $\sim 35$\%\ in H$\alpha$) during
the $\sim 20$ year period covered by the HST obserrvations.

\section{The [O~III] $\lambda$ 5007 images}

There are also two [O~III]~$\lambda$~5007 HST images of HH~1/2 (see Table 1), with a time separation of
$\approx 20$~yr. These two images are shown in Figure 11, together with [O~III]~5007/H$\alpha$
line ratio maps for the two epochs. The line ratios have been computed only for the
regions with an H$\alpha$ intensity larger than $3\times 10^{-15}$~erg~s$^{-1}$~cm$^{-2}$~arcsec$^{-2}$.

It is clear that the morphology of the [O~III]~5007 emission has changed rather dramatically
over the past 20 years. It has become angularly more extended, and has developed a high
[O~III] 5007/H$\alpha$ region along the W wing of HH~1F. Also, in both epochs we see a
central, SE to NW filament with a relatively high [O~III]/H$\alpha\approx 0.2$ value,
along the outflow axis. This region coincides spatially with the ``high proper motion
velocity channel'' discussed in section 4.

The [O~III] flux (integrated over the $18\times 18$~arcsec field shown in Figure 11) has
a substantial decrease of a factor $\sim 2$ from 1994 to 2014 (see the last column
of Table 2). The [O~III]/H$\alpha$ line ratio of the HH~1 field has a value of
$\approx 0.48$ in 1994 and $\approx 0.25$ in 2014.

\section{Summary: the time evolution of HH~1}

As described in sections 3-5, an analysis of four epochs of H$\alpha$ and [S~II] HST images
(covering a time span of 20 years) gives somewhat surprising results. The proper motions
of HH~1 have been measured for many decades:
\begin{itemize}
\item Herbig \& Jones (1981) measured proper motion velocities of HH~1,
  obtaining a velocity of 320~km~s$^{-1}$ for HH~1F and lower velocities (down to 143~km~s$^{-1}$
  for HH~1A) for the trailing condensations from plates obtained from 1959 to 1980. They
  also determined a velocity of 444~km~s$^{-1}$ from an earlier series of plates (obtained
  from 1946 to 1980) in which HH~1 is unresolved. These authors attribute this larger velocity
  to the fact that the morphology of HH~1 changed rather dramatically during the 1946-1980
  period, as in the earlier images the emission was dominated by the region around the HH~1A
  condensation,
\item Eisl\"offel et al. (1994) analyzed a set of [S~II] CCD images covering the 1986-1994
  period, and found a proper motion velocity of 351~km~s$^{-1}$ for HH~1F, and of 98~km~s$^{-1}$
  for HH~1A,
\item Bally et al. (2002) used HST images in 1994.61 and 1997.58 to obtain peak proper motion
  velocities for HH~1F of 320~km~s$^{-1}$ (for [S~II]) and 368~km~s$^{-1}$ (for H$\alpha$). These
  proper motions are consistent, but not identical (see Table 2) to the ones that we are now
  determining from these two frames (due to the differences in the techniques with which they
  have been determined).
\end{itemize}
The velocities given in the items above have been renormalized to a distance of 414~pc to the
HH~1/2 system.

From this, we see that the proper motion velocities observed for the leading region of HH~1 (HH~1F)
appear to have been stable, with values of $\sim 350$~km~s$^{-1}$ from 1959 to 1997. From our analysis
of the two later sets of HH~1/2 images obtained with the HST (epochs 2007.63 and 2014.63, see Table 1)
we find that in more recent times HH~1F appears to have increased its velocity. This is seen by the
fact that the maximum axial velocity of HH~1F has increased from $\sim 300$ to $\sim 600$~km~s$^{-1}$
in H$\alpha$ and from $\sim 330$ to $\sim 500$~km~s$^{-1}$ in [S~II] over the three frame pairs
obtained from 1994 to 2014 (see Table 2). During this period, we also see an increase in the proper
motion velocities obtained for a cross-correlation box which includes the region of the head of HH~1
(see Figures 1 and 2). The acceleration of HH~1 is also seen in the proper motion maps (Figures 5 and 7)
and velocity distributions (Figure 8) obtained from a study of the spatially resolved proper motions of the
H$\alpha$ and [S~II] condensations.

Interestingly, the decrease in the [S~II] emission of HH~1 observed by Eisl\"offel et al. (1994) over
the $1987\to 1994$ period (see their Figure 4) appears to have continued only until 2007,
and the [S~II] emission has a higher value in 2014 (see Figure 9).
The H$\alpha$ emission also decreased from
$1994\to 2007$, and shows an increase in the 2014 HST image. The [S~II]/H$\alpha$ ratio
first decreases from 0.79 (in 1994) to 0.76 (in 1997), and then grows monotonically, attaining
a value of 0.99 in 2014.

The fact that the minimum in the line ratio does not coincide (temporally)
with the minima of the line fluxes (see Figure 9) rules out a simple explanation
of the line variability as a result of variations of the foreground extinction. Also,
a variability of the extinction would produce higher [S~II]/H$\alpha$ ratios at
the times of higher extinction (i.e., of lower observed line intensities). Such
a correlation is clearly not seen in HH~1 (see Figure 9)

Finally, we see that in recent years the size of HH~1 (along the outflow axis) has increased
quite considerablly. The most dramatic increase is obtained for the H$\alpha$ emission, which
has grown in extent from $\sim 10$ to $\sim 14$~arcsec from 1994 to 2014 (see Figure 10). The morphology
observed in the two available [O~III] images (1994 and 2014, see Table 1) also presents a large
variability (see Figure 11) and the [O~III]/H$\alpha$ ratio (of the line fluxes integrated
over the emitting region of HH~1) has a decrease from 0.48 (in 1994) to 0.25 (in 2014).

It therefore appears that HH~1 is presently going through a period of substantial
changes in morphology, and in the H$\alpha$ and [S~II] intensities and ratios. These changes are occurring
over timescales of a few years, as evidenced by the large differences observed between the
2007 and 2014 observations (see Figures 9 and 10). These changes in the emission are accompanied
by the appearance of larger proper motions in some of the features of HH~1 (see Figures 5, 7 and 8).

Given the fact that the proper motion velocities of HH~1 appear to
be stable over relatively long timescales (from 1959 to 1987, see above), we speculate that
the present period of apparent acceleration is likely to be an event of relatively
short duration. In principle, we could be seeing an event in which an ``internal working
surface'' (produced by an ejection velocity variability, see, e.g., Raga et al. 1990b) is catching
up with the head of the jet. This would lead to a momentary increase in the speed of the jet
head (as modelled by Cant\'o \& Raga 2003 and Raga \& Cant\'o 2003).
Such a catching-up would be a complicated event in a flow with precession or other
direction variability (see, e.g., Raga et al. 2010), and deserves an exploration
in terms of 3D numerical simulations. This would illustrate whether or not such catching-up
events can reproduce the present time-evolution of HH~1.

Alternatively, HH~1 might be travelling into an environment with decreasing densities at larger
distances from the outflow source. Such a scenario would be consistent with the model of
Henney (1996), who suggested that the side-to-side asymmetries of HH~1 might be the
result of an environmental density gradient $perpendicular$ to the outflow axis. An exploration
of this scenario with numerical simulations would also be very interesting.

The present period of large variability might resemble the one reported by Herbig (1973), who found
that HH~1F increased dramatically in brightness (to become the dominant knot of HH~1) within
a gap in a series of Lick plates between 1968 and 1973. It is a real pity that the details of
this fast, 1968 to 1973 evolution have not been observed. It appears that we now have a second
chance of observing major changes in HH~1, and it would definitely be worthwhile to
obtain (at least reasonably high angular resolution) images in a few lines
and with a good time coverage of HH~1 during the following few years.

We should point out that in measuring proper motions, relative changes in the intensites of
different regions of an HH object can lead to proper motion velocities which do not correspond
to a real displacement of an emission structure (as described in Section 1, this was an initial
worry of Herbig, who later concluded that real motions were indeed detected). However, for
knot structures that can be recognized (from their similar intensities and morphologies) in
pairs of epochs, showing displacements that are larger than the sizes of the knots, there
is little doubt that the knots indeed have coherent motions in the plane of the sky.

The observed proper motions of course could be motions of the gas itself (``matter motions'') or a
wave structure travelling through a gas with a different motion. However, in a hypersonic flow such
as a jet from a young star (with sonic and Alv\`enic
Mach numbers of $\sim 10\to 100$, deduced from radial velocities,
proper motions and plasma diagnostic determinations of the temperature), waves travelling through
the flow are highly likely to be shock waves, which (for the densities and velocities measured
in HH objects) are highly radiative. Such radiative shocks have very high compression ratios, so that
the post-shock gas has a motion which closely follows the ``normal motion'' of the shock wave (i.e.,
the motion perpendicular to the shock front). Therefore, the motions measured for the emitting
gas behind a radiative shock wave closely follow the ``matter motion''.

Having said this, it is also clear that in measuring proper motions of details seen in high
resolution images of HH objects (such as the ones described in the present paper), at least
in some cases one might be confusing relative brightness changes with actual motions of coherent structures.
This is an unavoidable problem of carrying out proper motion measurements of extended structures
in the ISM.

However, this is not a major problem for carrying out comparisons with numerical
simulations of different scenarios for producing the flow. For example, one can compare the
present proper motion measurements of HH~1 with numerical simulations (e.g., of a radiative jet
from a variable source) in which one predicts the time-evolution of emission line intensity
maps. This can be done, e.g., applying the proper motion detection algorithm described in
Section 4 to a time-sequence of predicted maps, and comparing the proper motions obtained
in this way with our measurements of HH~1 proper motions. The possible confusions
between ``matter motions'' and relative intensity changes (if present) should occur
in the analyses of both the observed and the numerically predicted emission line maps,
so that a comparison between observations and models is in principle valid.

\begin{deluxetable}{lllc}
\tablecaption{HST images of the HH~1/2 system\label{obs}}
\tablewidth{0pt}
\tablecolumns{9}
\tablehead{
  \colhead{Epoch} & \colhead{filters} & \colhead{emission lines} & \colhead{exposures [s]}}
\startdata
1994.61  &  F502N & [O III] 5007 &  3000  \\
         &  F656N & H$\alpha$    &  3000  \\
         &  F673N & [S II] 6716/6731 & 3000 \\
1997.58  &  F656N & H$\alpha$    &  2000  \\
         &  F673N & [S II] 6716/6731 & 2200 \\
2007.63  &  F656N & H$\alpha$    &  2000  \\
         &  F673N & [S II] 6716/6731 & 1800 \\
2014.63  &  F502N & [O III] 5007 &  2798  \\
         &  F656N & H$\alpha$    &  2686  \\
         &  F673N & [S II] 6716/6731 & 2798 \\
\enddata
\end{deluxetable}

\begin{deluxetable}{cccccccccc}
\tablecaption{H$\alpha$ and [S~II] velocities and fluxes\label{hasii}}
\tablewidth{0pt}
\tablecolumns{10}
\tablehead{
\colhead{Epoch$^a$} & \colhead{$v_\parallel(H\alpha)$$^b$} & \colhead{$v_\perp(H\alpha)$$^b$} &
  \colhead{$v_\parallel([S\,II])$$^b$}  & \colhead{$v_\perp([S\,II])$$^b$} &
  \colhead{$\Delta v_\parallel$$^c$} & \colhead{$\Delta v_\perp$$^c$} & \colhead{$F_{H\alpha}$$^d$} &
      \colhead{$F_{[S\,II]}$$^d$} &  \colhead{$F_{[O\,III]}$$^d$} \\
  & \multispan6{\hfill[km~s$^{-1}$]\hfill} &
\multispan3{\hfil{[$10^{-13}$ erg s$^{-1}$ cm$^{-2}$]}\hfil}}
\startdata
1994.61  &           &        &              &                   &               &            &  7.79 & 6.14 & 3.74 \\
(1996.1)  &  (0.8,  & ($-$173.4, &  (2.4,  & ($-$137.6, &  381.9 &  206.6  &  &  \\
          &  297.8) & $-$47.9)   &  382.7) & $-$2.7)    &         &        &  &  & \\
1997.58  &           &        &              &                   &               &            &  6.87 & 5.21 & \\
(2002.6)  &  (90.3, & ($-$80.5,  &  (3.6, & ($-$147.8, &  305.8 &  227.0  &   &  \\
          &  273.0) &    52.6)   &  309.4) & 79.2)      &         &        &  &  & \\
2007.63  &           &        &              &                   &               &            &  5.70 & 5.04 & \\
(2011.1)  &  (61.7,) &  ($-$166.7, &  (42.8, & ($-$104.0, &  557.3 &  236.7  &  &  \\
          & 600.1    & 70.0)       &  358.9) &  $-$27.7)     &         &        &  & &  \\
2014.63  &           &        &              &                   &               &            &  7.76 & 7.68 & 1.97 \\
\enddata
{\baselineskip=0pt
  \tablenotetext{a}{The four observed epochs are listed, as well as the average time corresponding to the
    successive pairs of epochs (in parentheses)}
  \tablenotetext{b}{The pairs of numbers in parentheses correspond to the minimum and maximum values of
    the proper motion velocities along and across the outflow axis}
  \tablenotetext{c}{The values of $\Delta v_\parallel$ and $\Delta v_\perp$ correspond to the whole range
    covered by both the H$\alpha$ and [S~II] proper motions}
  \tablenotetext{d}{The fluxes correspond to all of the emitting region of HH~1}
}
\end{deluxetable}

\begin{figure}
\centering
\includegraphics[width=9cm]{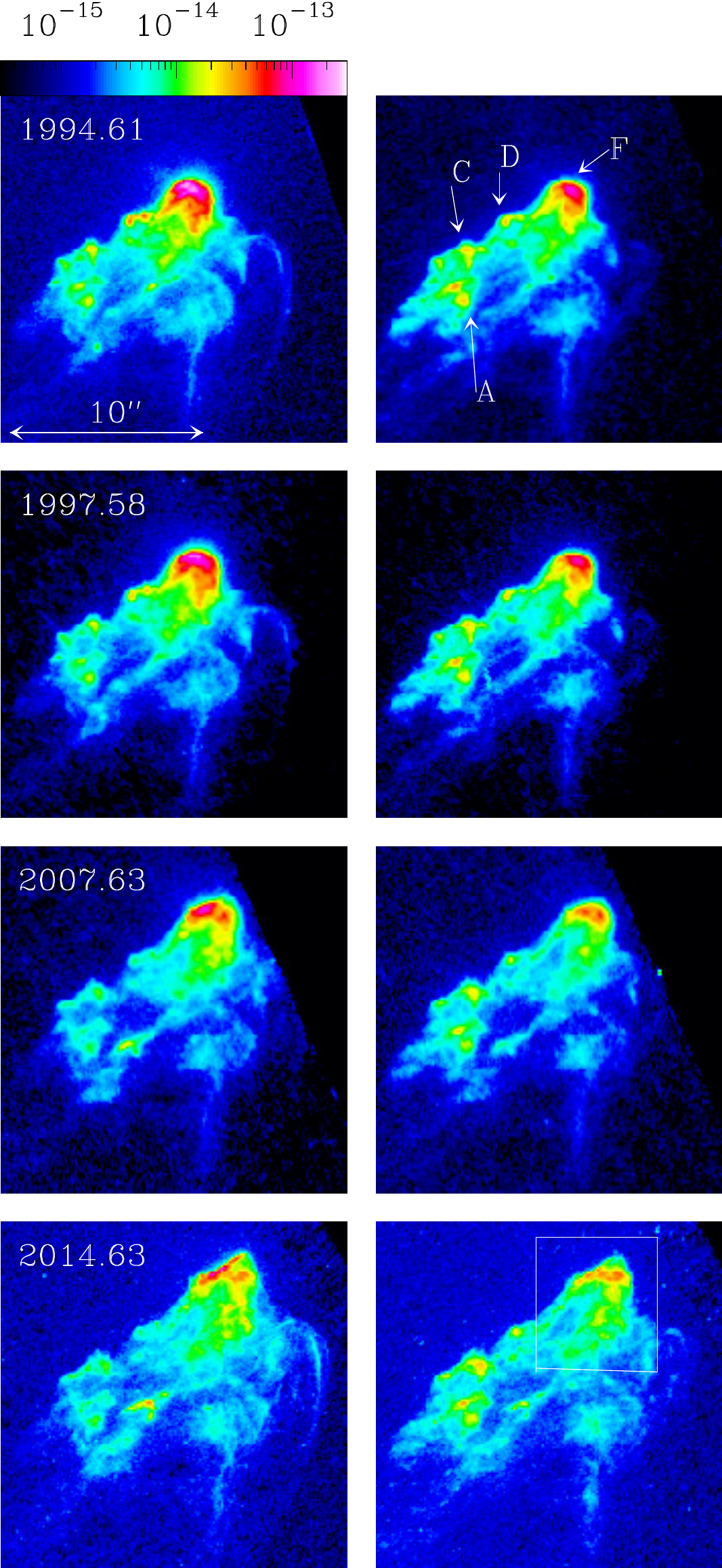}
\caption{H$\alpha$ (left) and [S~II]~$\lambda\lambda$~6716/6731 (right)
  images of the four available epochs of HST observations (the
  times of the observations are given by the labels on the top left of the H$\alpha$ maps). The
  displayed field includes all of the emitting region of HH~1, and the emission is shown with the logarithmic
  colour scale given (in erg~s$^{-1}$~cm$^{-2}$~arcsec$^{-2}$) by the top left bar. The
  identifications of the emitting regions of HH~1 (following Herbig \& Jones 1981) are shown in the top right frame,
  and the angular scale of the maps is shown in the top left frame.
  The white box in the bottom right frame shows the region used for the cross correlations discussed in section 3.}
\end{figure}

\begin{figure}
\centering
\includegraphics[width=10cm]{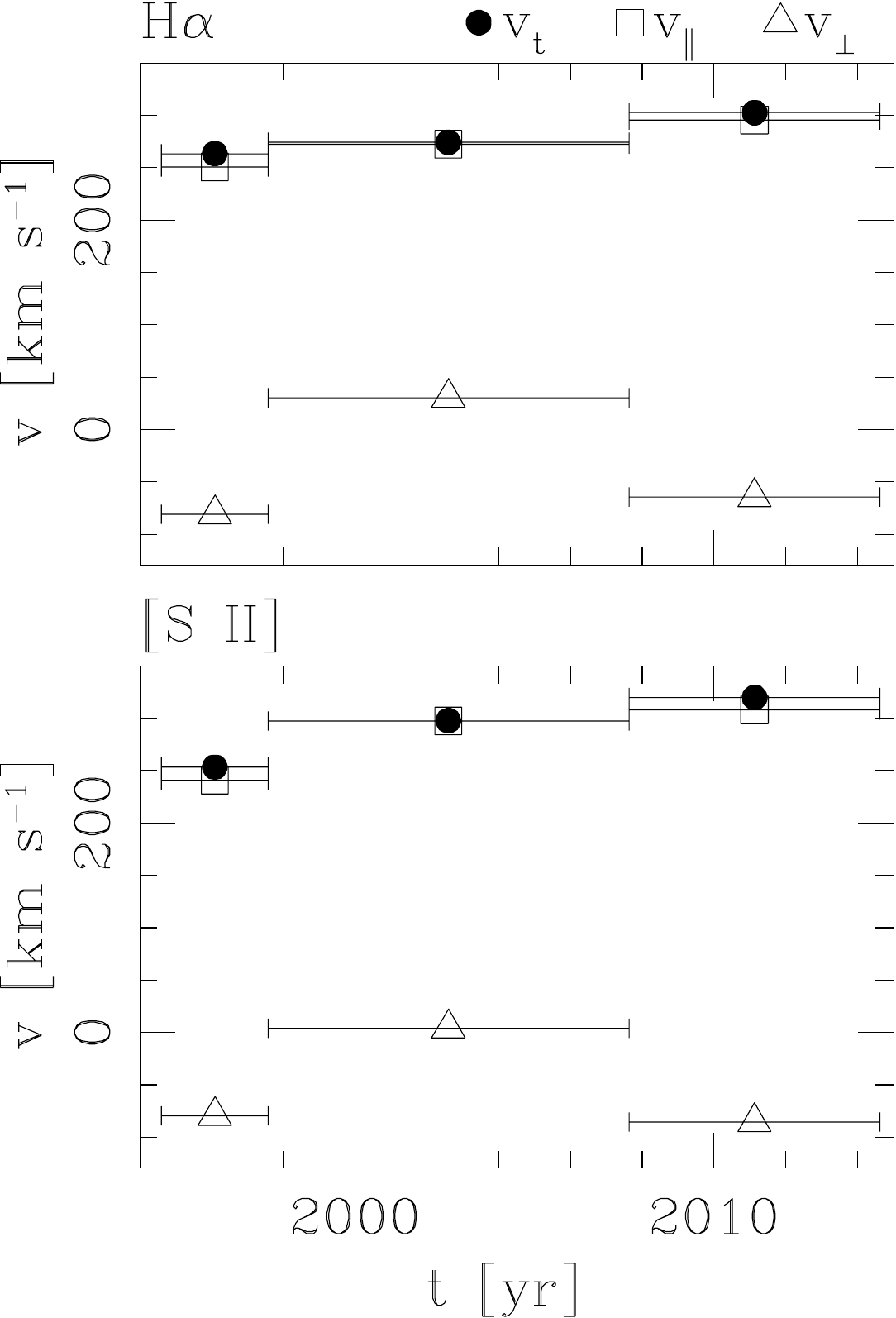}
\caption{Cross-correlation proper motion velocities of HH~1 (see section 3) for the three pairs of HST image
  epochs. The modulus of the proper motion velocity (filled circles), and velocities parallel (empty squares),
  and perpendicular (empty triangles) obtained from the H$\alpha$ (top) and [S~II]~$\lambda\lambda$~6716/6731
  frames (bottom)
  are shown. The horizontal bars represent the intervals between the successive epochs, and the points are
  centered at the average between the times at which each pair of successive frames were obtained. The formal
  errors of the proper motion velocities are of order of $\sim 5$~km~s$^{-1}$. In these plots, $v_\parallel$ is
  the proper motion velocity along and $v_\perp$ the velocity across the
  outflow axis. Positive values of $v_\perp$ are directed to the NE, and negative values to the SW. The
  total proper motion velocity is $v_t=\sqrt{v_\parallel^2+v_\perp^2}$ (also shown in the two frames).} 
\end{figure}

\begin{figure}
\centering
\includegraphics[width=8cm]{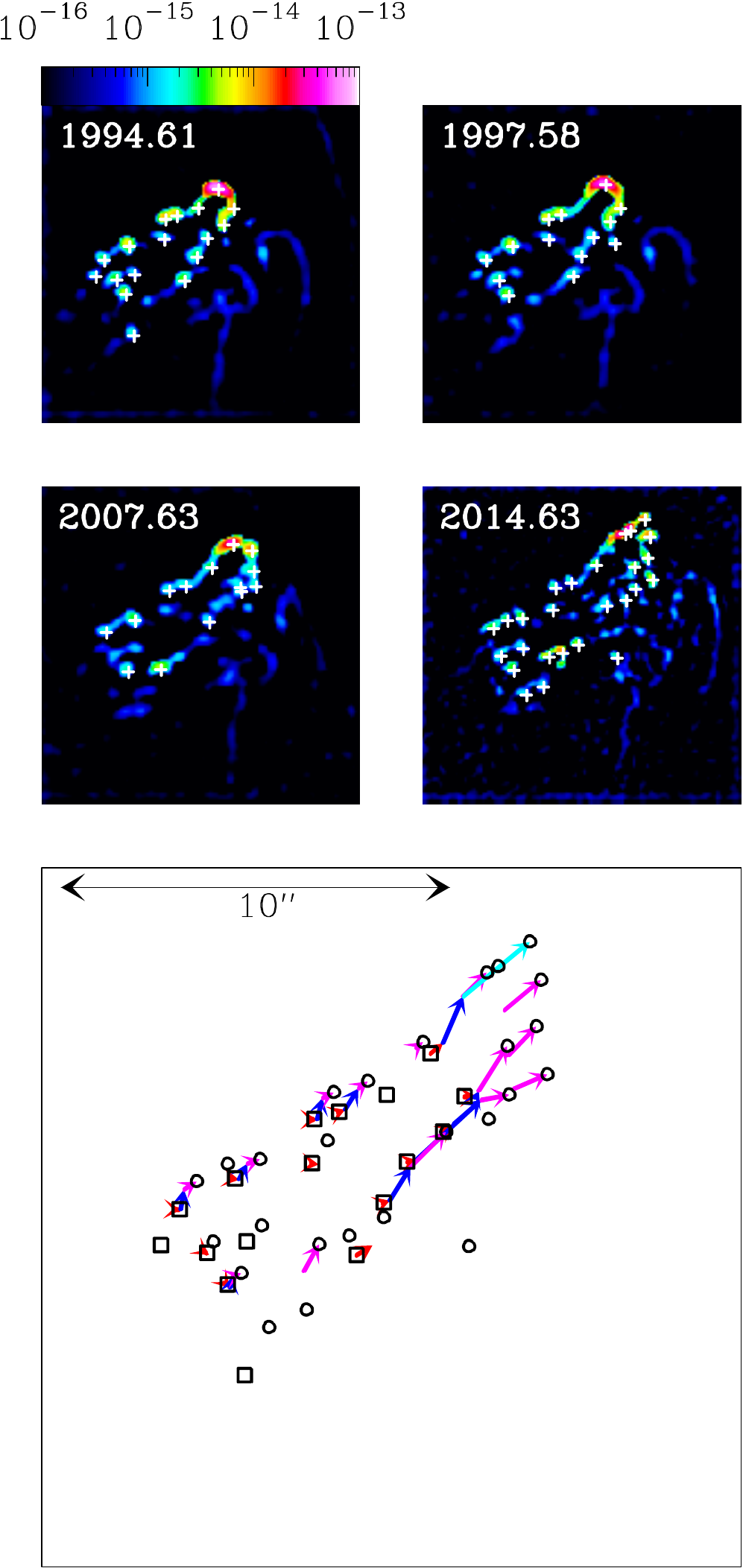}
\caption{The four top frames (labeled with the times at which the observations were made) show the H$\alpha$ maps
  convolved with a $\sigma=3$~pix ($0.3$~arcsec) wavelet (see section 4 and equation 1). The convolutions are shown with
  the logarithmic scale given (in erg~s$^{-1}$~cm$^{-2}$~arcsec$^{-2}$) by the top left bar. The displayed field is identical
  to the one shown in Figure 1. The white crosses show the positions of the peaks identified in each of
  the convolved images (see section 4). The bottom frame (which includes the angular scale)
  shows a blow-up of the same field, with
  the intensity peaks found in the 1994.61 (squares) and 2014.63 (circles) frames. The arrows show the angular
  offsets between identified pairs in the (1994.61, 1997.58) epochs (red arrows), in the (1997.58, 2007.63) pair
  (dark blue arrows) and (2007.63, 2014.63) pair (purple arrows). The cyan arrow corresponds to the motion
  determined for the leading condensation of HH~1F (done with the backwards companion search described in
  section 4) in the (2007.63, 2014.63) pair of convolved images.}
\end{figure}

\begin{figure}
\centering
\includegraphics[width=8cm]{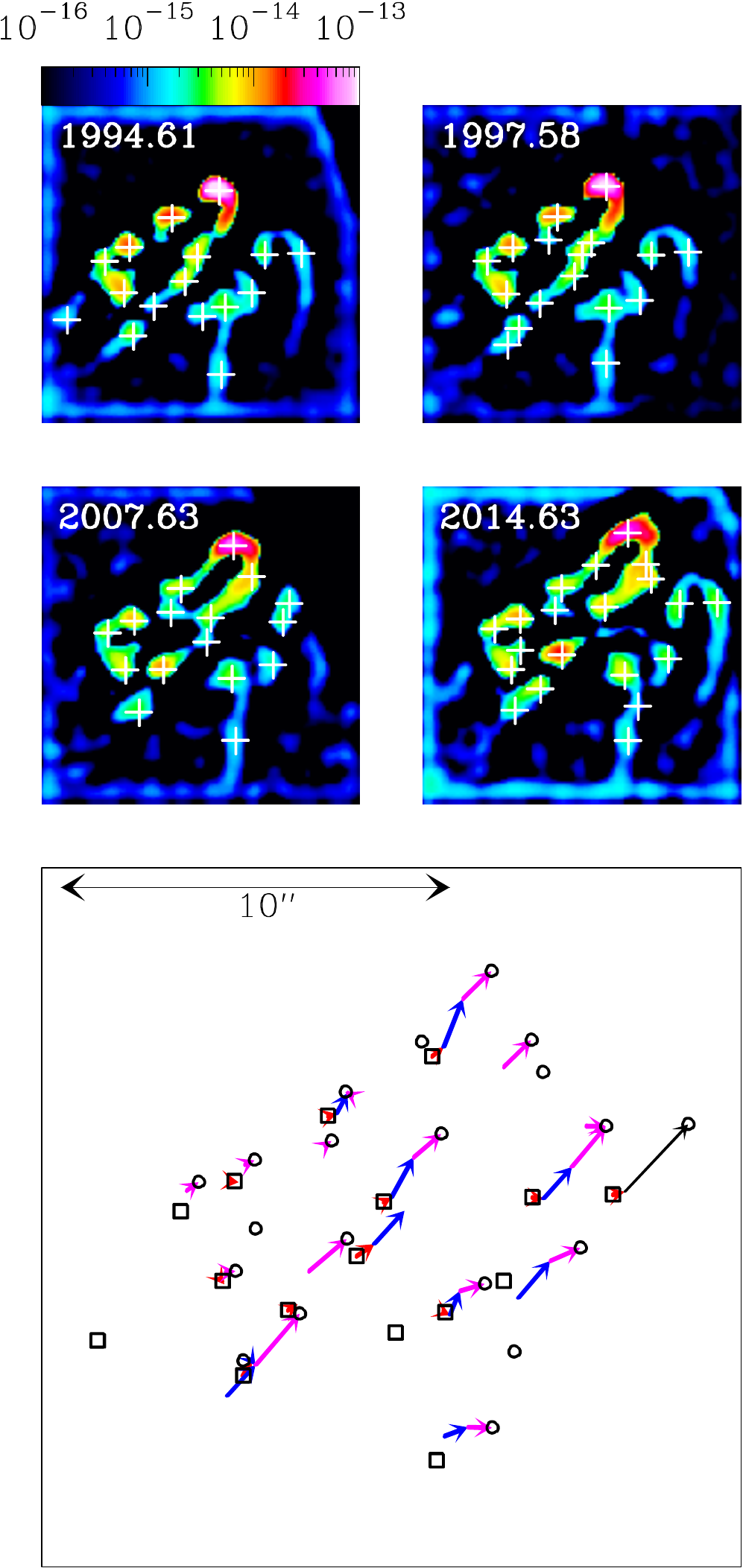}
\caption{The same as Figure 3, but for H$\alpha$ images convolved with a $\sigma=5$~pix ($0.5$ arcsec) wavelet. In the
  bottom frame, the red arrows represent the offsets of identified knot pairs in the (1994.61, 1997.58) epochs, the
  blue arrows for the (1994.61, 1997.58) epochs and the purple arrows for the (2007.63, 2014.63) epochs. The black
  arrow shows an offset obtained from the (1997.58, 2014.63) pair of frames (as this region lies outside the
``HST footprint'' in the 2007.63 frame, see section 4).} 
\end{figure}

\begin{figure}
\centering
\includegraphics[width=8cm]{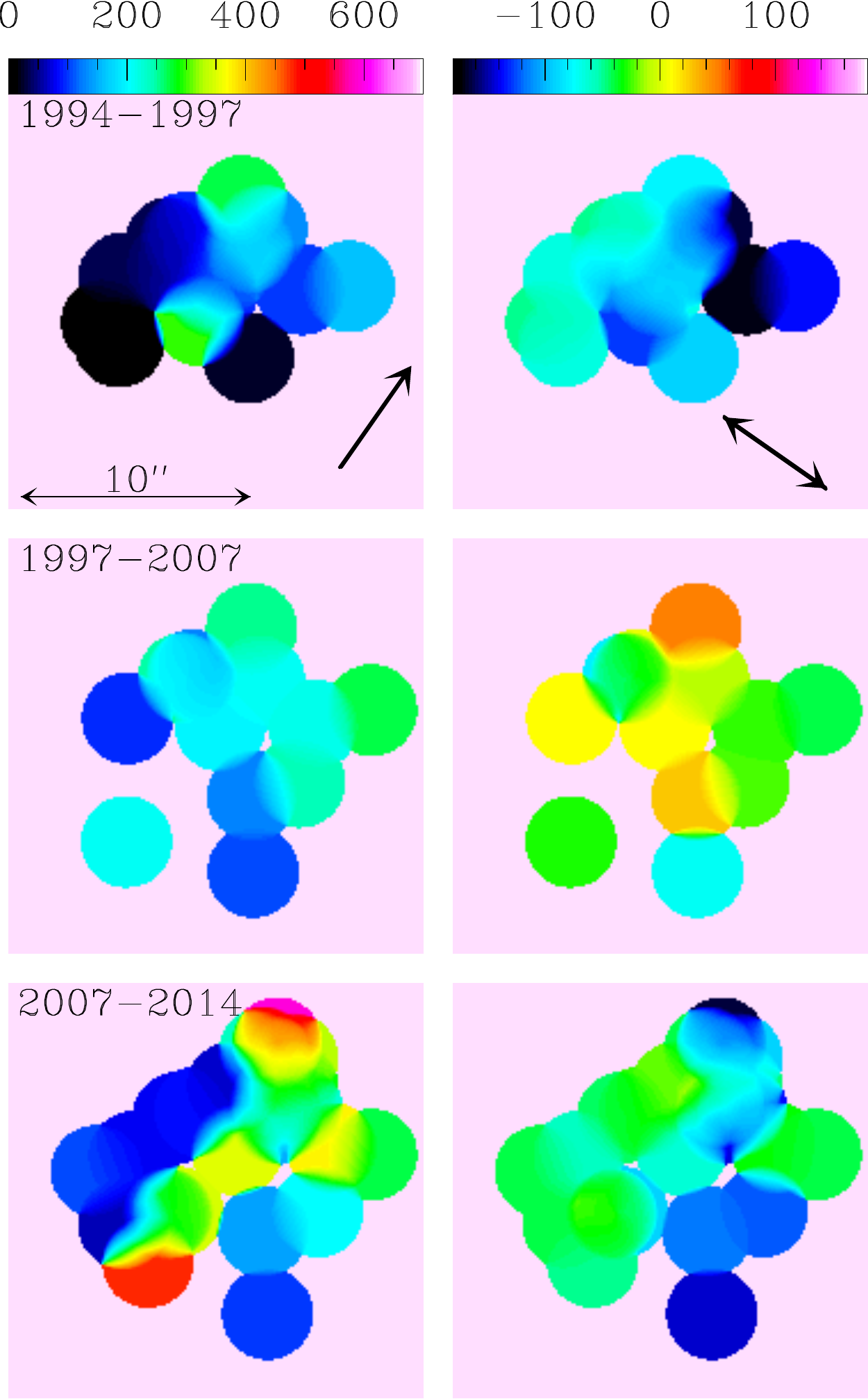}
\caption{Maps of the H$\alpha$ proper motion velocities parallel (left) and perpendicular (right) to the outflow axis
  for the three pairs of consecutive epochs. The directions
  of the velocities are shown with the thick black arrows in the top frames. Positive velocities perpendicular to the axis
  are directed to the NE. The frames are labeled with the times of the two epochs used to determine
  each set of proper motion velocities.}
\end{figure}

\begin{figure}
\centering
\includegraphics[width=8cm]{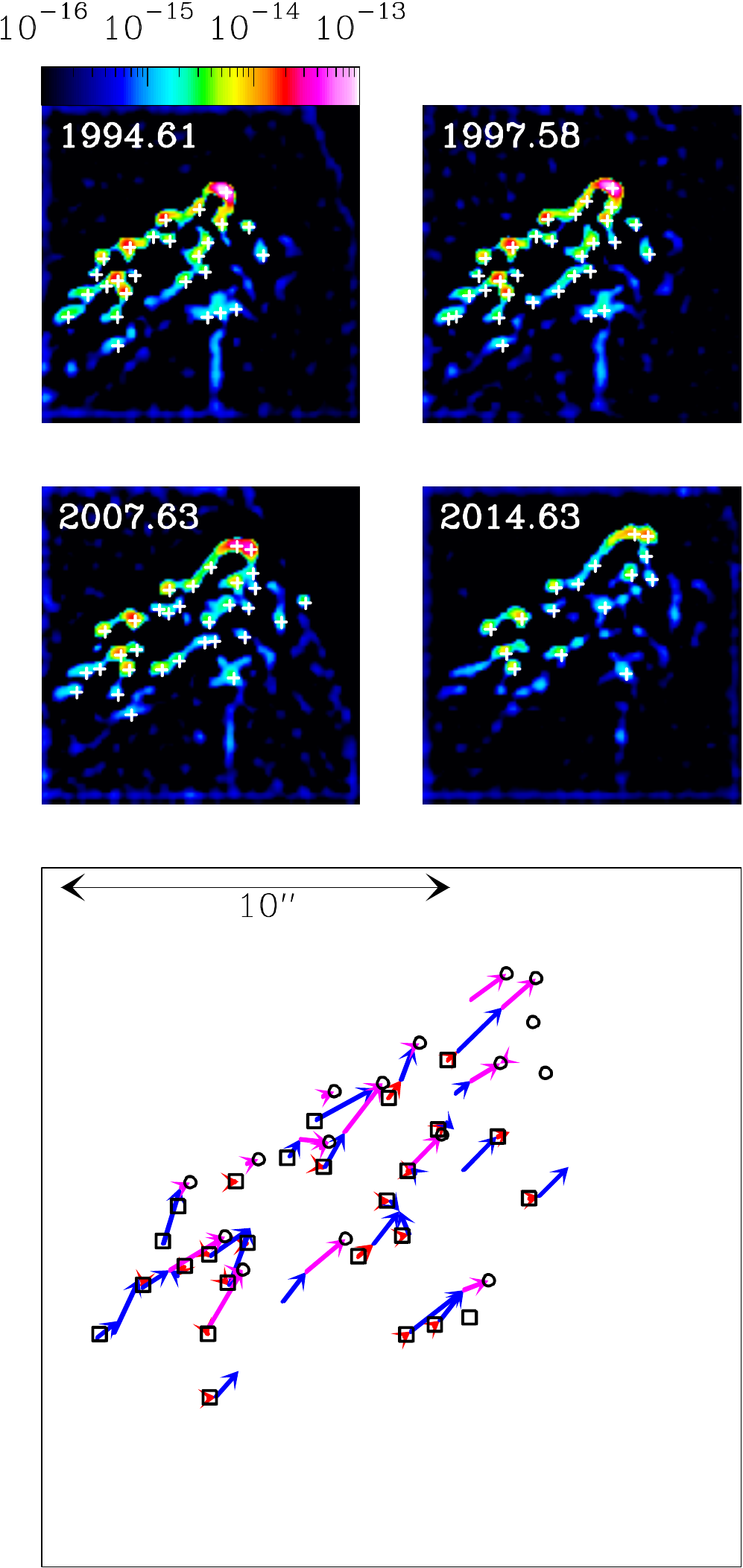}
\caption{The same as Figure 2, but for the [S~II]~$\lambda\lambda$~6716/6731 images convolved with a $\sigma=3$~pix wavelet.} 
\end{figure}

\begin{figure}
\centering
\includegraphics[width=8cm]{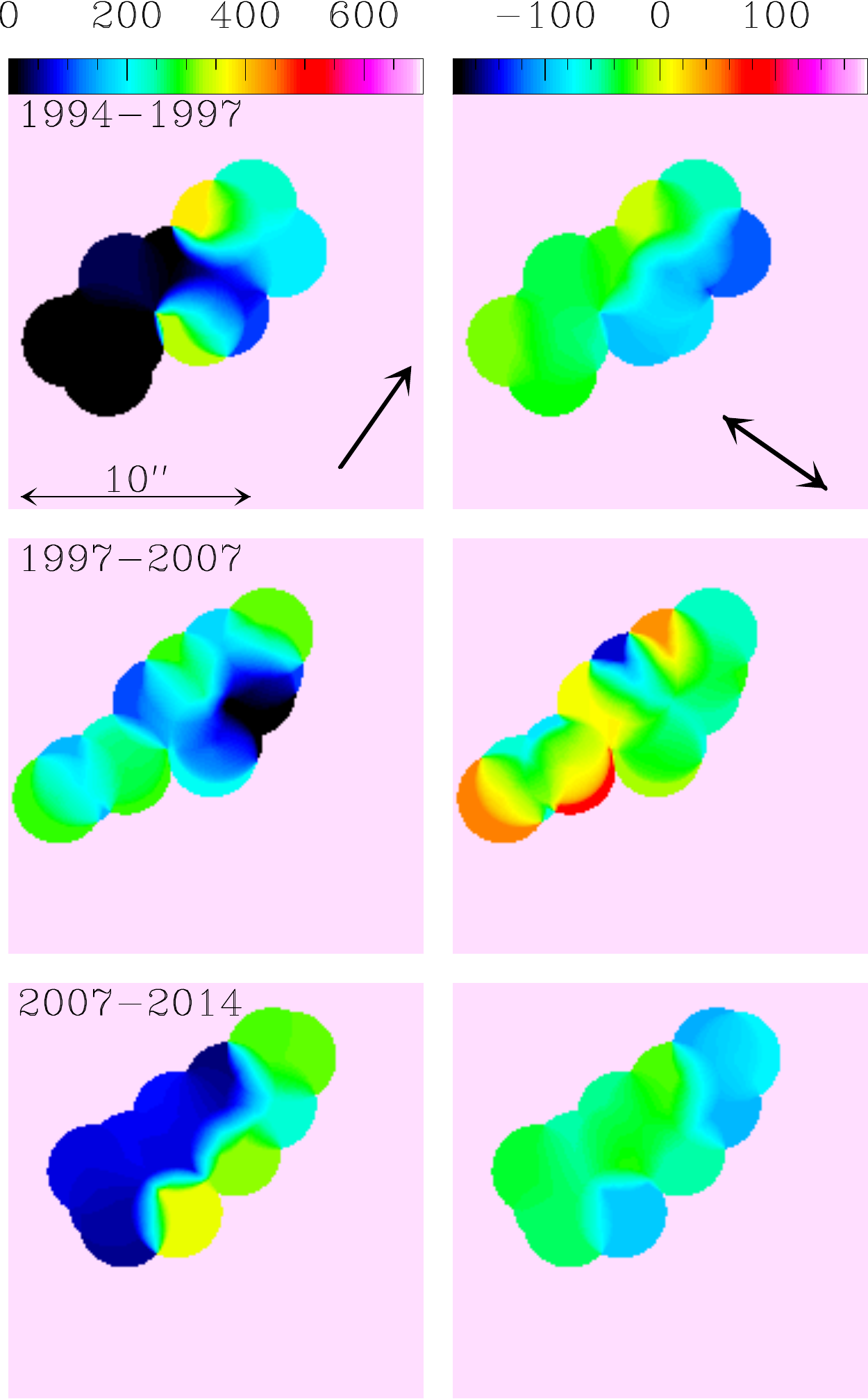}
\caption{Maps of the [S~II]~$\lambda\lambda$~6716/6731 proper motion velocities parallel
  (left) and perpendicular (right) to the outflow axis
  for the three pairs of consecutive epochs (each frame labeled with the times of the two epochs). The directions
  of the velocities are shown by the thick black arrows in the top frames. Positive velocities perpendicular to the axis
  are directed to the NE.}
\end{figure}

\begin{figure}
\centering
\includegraphics[width=10cm]{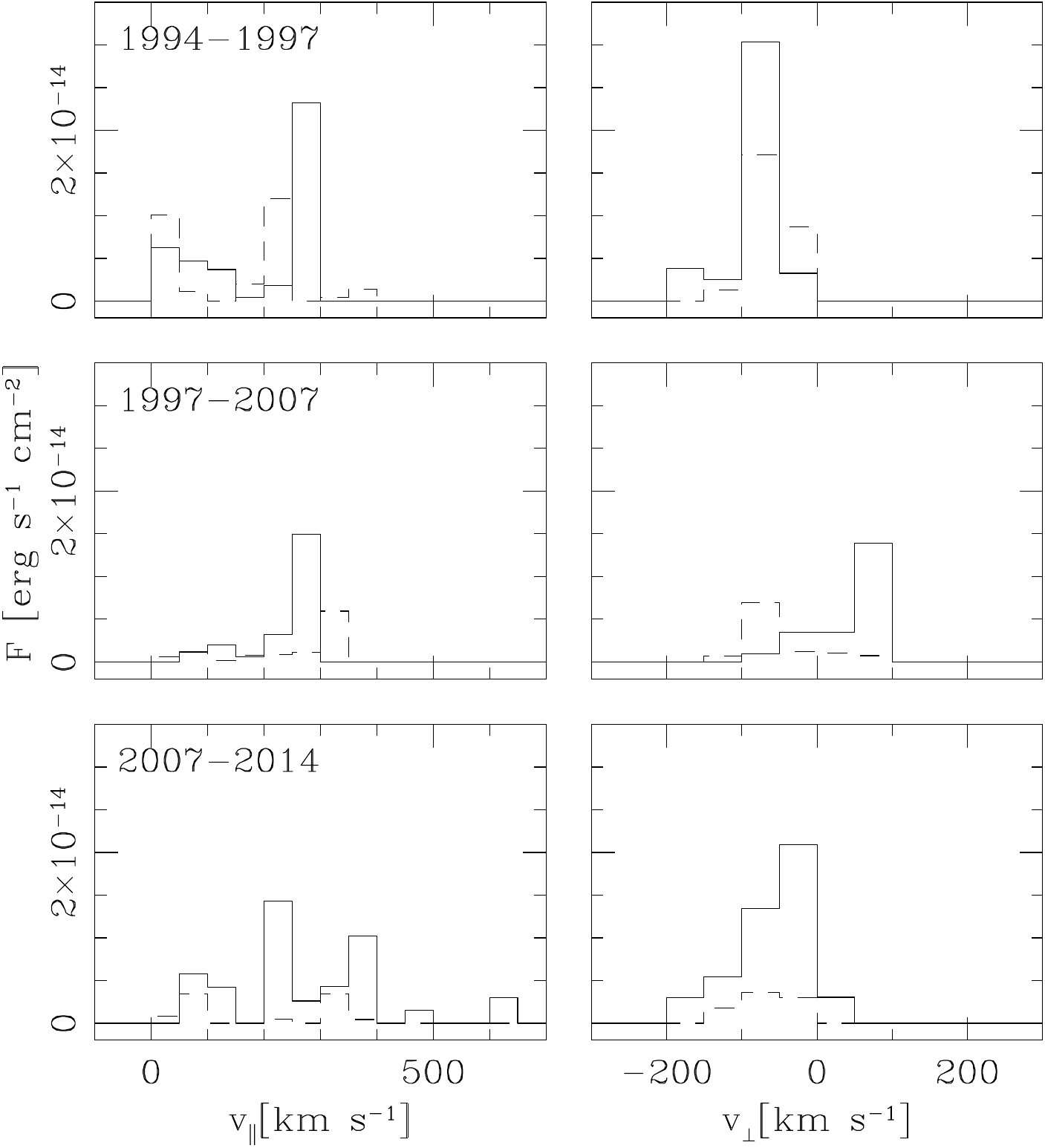}
\caption{
  The intensities of the knots (in the second frame of each pair of epochs) added over bins
  of $\Delta v=50$~km~s$^{-1}$ of proper motion velocities parallel (left) and perpendicular (right) to the outflow axis,
  for the three pairs of epochs. The frames are labeled with the times of the pairs of successive epochs.
  The solid histograms show the H$\alpha$ and the dashed histograms the [S~II] distributions.}
\end{figure}

\begin{figure}
\centering
\includegraphics[width=10cm]{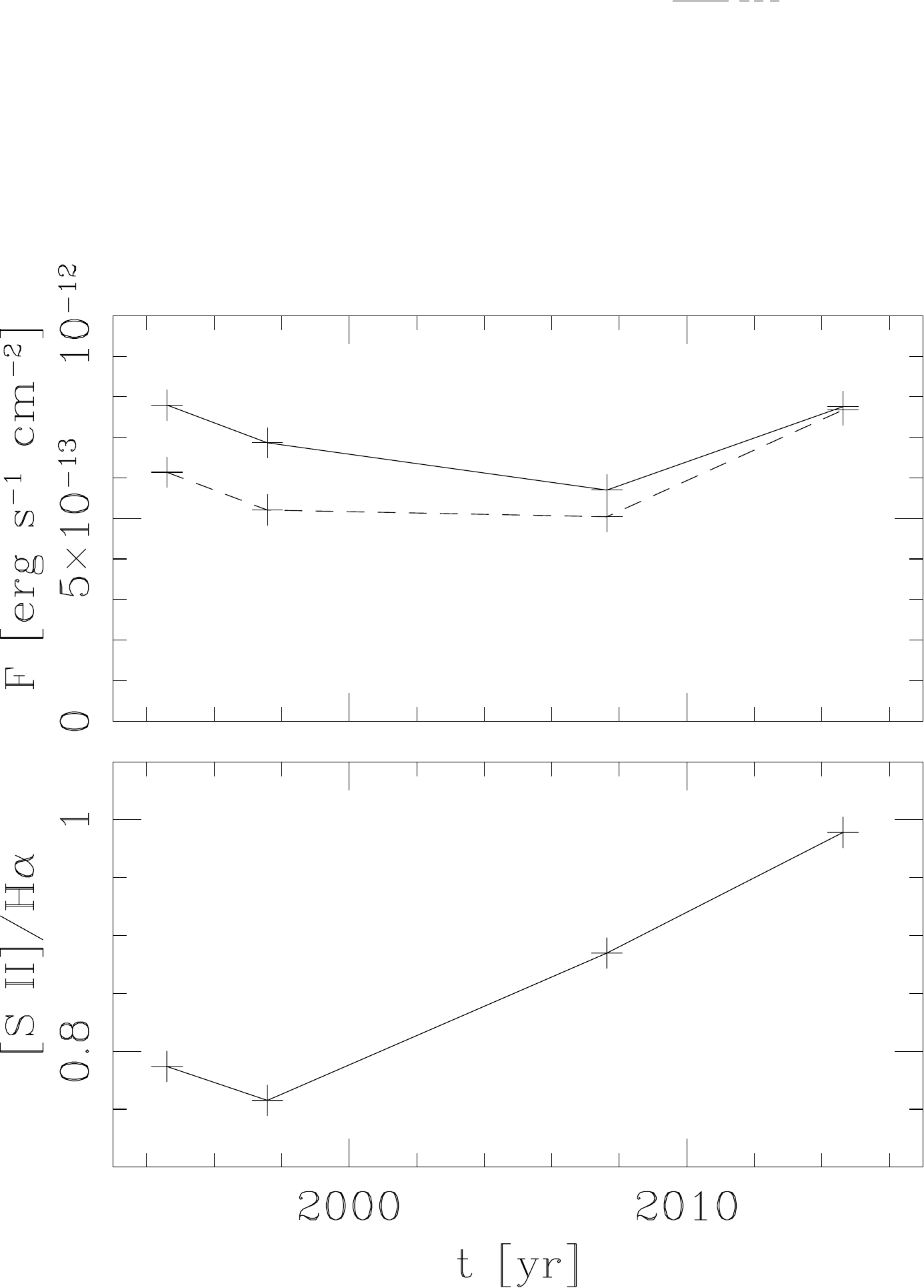}
\caption{The top frame shows the H$\alpha$ (solid line) and [S~II]~$\lambda\lambda$~6716/6731 (dashed line)
  fluxes of the whole emitting region of HH~1 (defined as the $18\times 18$~arcsec field shown in Figure 1)
  in the four epochs of HST images. The bottom frame shows the resulting [S~II]/H$\alpha$
  line ratio.}
\end{figure}

\begin{figure}
\centering
\includegraphics[width=10cm]{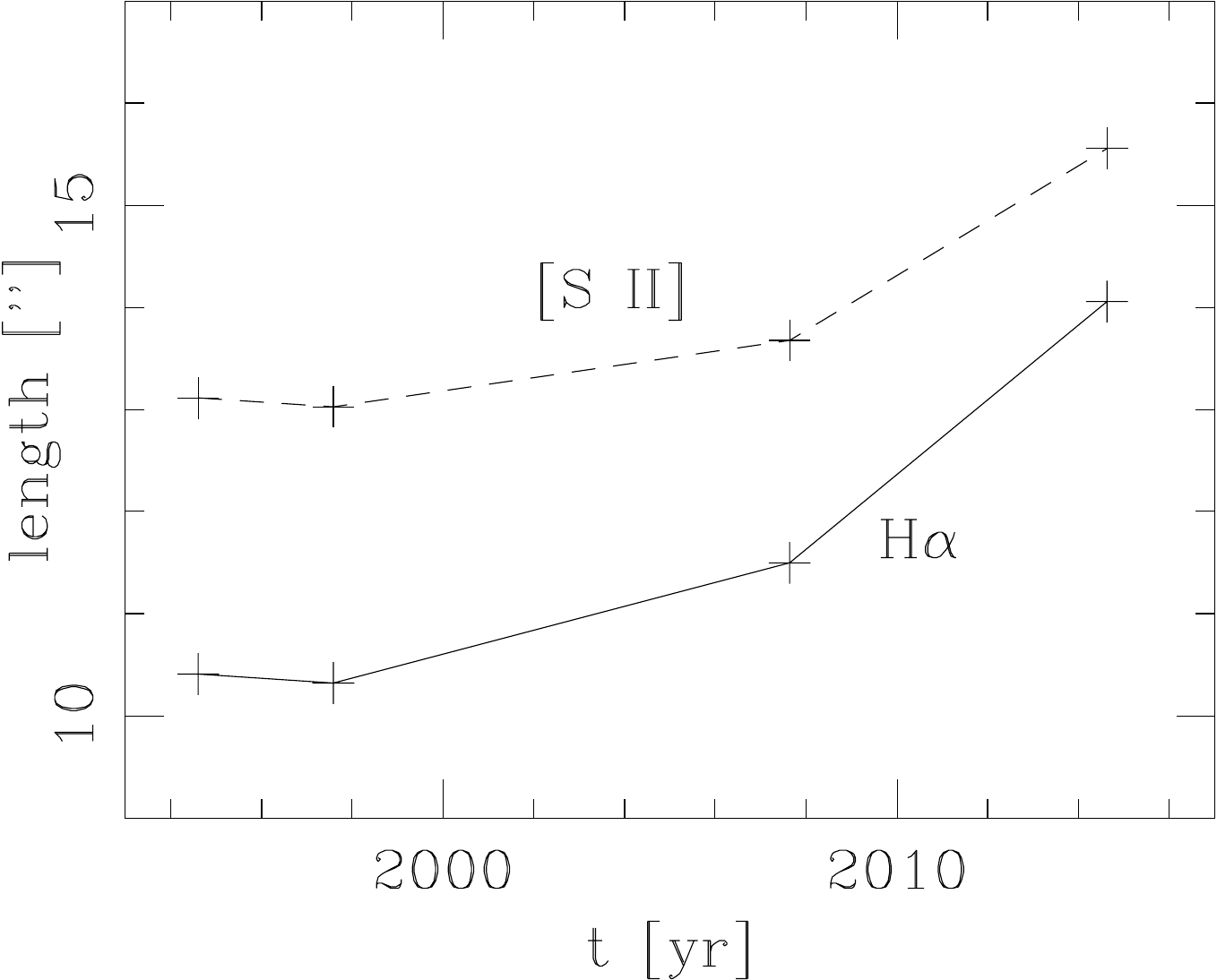}
\caption{Size along the outflow axis of the $3\times 10^{-15}$~erg~cm$^{-2}$s$^{-1}$~arcsec$^{-2}$ isophote obtained
from the H$\alpha$ (solid line) and [S~II]~$\lambda\lambda$~6716/6731 (dashed line) images of HH~1 as a function of time.}
\end{figure}

\begin{figure}
\centering
\includegraphics[width=12.5cm]{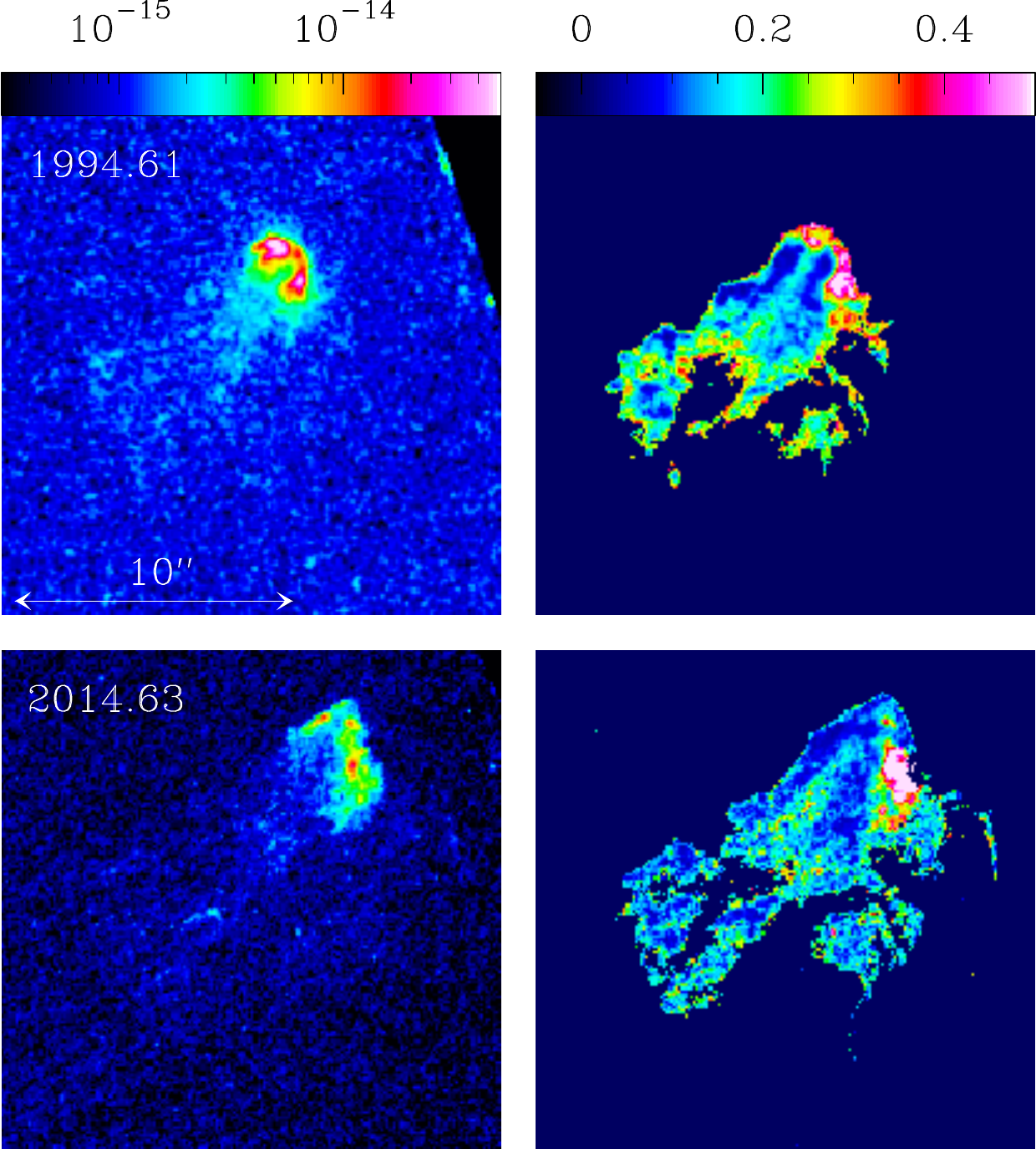}
\caption{The left column shows the two available epochs of [O~III]~$\lambda$~5007 images (shown with the logarithmic
  colour scale given in erg~s$^{-1}$~cm$^{-2}$~arcsec$^{-2}$ by the top bar). The right column shows the
  [O~III]~5007/H$\alpha$ ratio maps obtained for these two epochs. No reddening correction has been applied.}
\end{figure}

\begin{acknowledgements}
Support for this work was provided by NASA through grant HST-GO-13484 from the
Space Telescope Science Institute. AE and ACR acknowledge support from the CONACyT grants
101356, 101975 and 167611 and the DGAPA-UNAM grants IN105312, IN109715 and IG100214.
\end{acknowledgements}

%\clearpage


\begin{thebibliography}{}

\bibitem[{Bally et al.} (2000)]{bal02} Bally, J., Heathcote, S., Reipurth, B., Morse, J.,
%  Hartigan, P., \& Schwartz, R. 2002, AJ, 123, 2627

\bibitem[{B\"ohm et al.} (1994)]{boh93} B\"ohm, K. H., Noriega-Crespo, A.,
  \& Solf, J. 1994, ApJ, 416, 647

\bibitem[{Brugel et al.} (1985)]{bru85} Brugel, E. W., B\"ohm, K. H., B\"ohm-Vitense, E.,
\& Shull, J. M. 1985, ApJ, 292, L75

%\bibitem[B\"ohm \& Solf 1992]{boh92} B\"ohm, K. H., \& Solf, J. 1992, AJ, 104, 1193

\bibitem[{Brugel et al.} (1981a)]{bru81a} Brugel, E. W., B\"ohm, K. H., \& Mannery, E. 1981a, ApJS, 47, 117

\bibitem[{Brugel et al.} (1981b)]{bru81b} Brugel, E. W., B\"ohm, K. H., \& Mannery, E. 1981b, ApJ, 234, 874

%\bibitem[Haro 1952]{har52} Haro, G. 1952, ApJ, 115, 572


%\bibitem[Hartigan et al. 1987]{har87} Hartigan, P., Raymond, J. C., Hartmann, L. W.
%1987, ApJ, 316, 323

%\bibitem[Heathcote et al. 1996]{hea96} Heathcote, S. et al. 1996, AJ, 112, 1141

%\bibitem[Heng 2010]{hen10} Heng, K. 2010, PASA, 27, 23

%\bibitem[Herbig 1951]{her51} Herbig, G. H. 1951, ApJ, 113, 697

\bibitem[{Cant\'o \& Raga} (1994)]{can93} Cant\'o, J., \& Raga, A. C. 2003, RMxAA, 39, 261

\bibitem[{Cohen \& Schwartz} (1979)]{coh79} Cohen, M., \& Schwartz, R. D. 1979, ApJ, 233, L77

\bibitem[{Cudworth \& Herbig} (1979)]{cud79} Cudworth, K. M., \& Herbig, G. H. 1979, AJ, 84, 548

\bibitem[{Dudziak \& Walsh} (1997)]{dud97} Dudziak, G., \& Walsh, J. R., in {\it 1997 HST Calibration
  Workshop}, Space Telescope Science Institute, Eds. S. Casertano et al., 1997, p. 338

\bibitem[{Eisl\"offel et al.} (1994)]{eis94} Eisl\"offel, J., Mundt, R., \& B\"ohm, K. H.
  1994, AJ, 108, 1042

\bibitem[{Hartigan et al.} (2011)]{har11} Hartigan, P., et al. 2011, ApJ, 736, 29

\bibitem[{Heathcote \& Reipurth} (1992)]{hea92} Heathcote, S., \& Reipurth, B. 1992, AJ, 104, 2193

\bibitem[{Henney} (1996)]{hen96} Henney, W. J. RMxAA, 32, 3
  
\bibitem[{Herbig \& Jones} (1981)]{her81} Herbig, G. H., \& Jones, B. F. 1981, AJ, 86, 1232

\bibitem[{Herbig} (1968)]{her68} Herbig, G. H. 1968, in {\it Non-periodic phenomena
  in variable stars}, ed. L. Detre (Reidel), p.~75
  
\bibitem[{Herbig} (1973)]{her73} Herbig, G. H. 1973, in {\it Information Bulletin of
  Variable Stars}, 832

\bibitem[{Hester et al.} (1998)]{hes98} Hester, J. J., Stapelfeldt, K. R., \&
  Scowen, P. A. 1998, AJ, 116, 372

\bibitem[{Liseau et al.} (1996)]{lis96} Liseau, R., Huldtgren, M., Fridlund, C. V. M.,
  \& Cameron, M. 1996, A\&A, 306, 255

\bibitem[{Luyten} (1963)]{luy63} Luyten, W. J. 1963, Harvard Annu. Card No. 1589

%\bibitem[Miller 1968]{mil68} Miller, J. S. 1968, ApJ, 154, L57

%\bibitem[Noriega-Crespo \& Raga 2012]{nor12} Noriega-Crespo, A., \&
%Raga, A. C. 2012, ApJ, 750, 101

%\bibitem[Noriega-Crespo et al. 1997]{nor97} Noriega-Crespo, A.,
%Garnavich, P. M., Curiel, S., Raga, A. C., \& Ayala, S. 1997, ApJ, 486, L55

\bibitem[{O'Dell et al.} (2013)]{ode13} O'Dell, C. R., Ferland, G. J., Henney, W. J.,
Peimbert, M. 2013, AJ, 145, 92

%\bibitem[Ortolani \& D'Odorico 1980]{ort80} Ortolani, S., \& D'Odorico, S.
%1980, A\&A, 83, L8

%\bibitem[Pravdo et al. 1995]{pra95} Pravdo, S. H., Rodr\'\i guez, L. F.,
%Curiel, S., Cant\'o, J., Torrelles, J. M., Becker, R. H., \& Sellgren, K.
%1985, ApJ, 293, L35

%\bibitem[Pravdo et al. 2001]{pra01} Pravdo, S. H., Feigelson, E. D., Garmire, G.,
%Maeda, Y., Tsuboi, Y., \& Bally, J. 2001, Nature, 413, 708

%\bibitem[Raga et al. 2014]{rag14} Raga, A. C., Castellanos-Ram\'\i rez, A., Esquivel, A.,
%Rodr\'\i guez-Gonz\'alez, A., \& Vel\'azquez, P. F. 2014, RMxAA, submitted

\bibitem[{Raga et al.} (1990a)]{rag90a} Raga, A. C., Barnes, P. J., \& Mateo, M. 1990a,
AJ, 99, 1912

\bibitem[{Raga et al.} (1990b)]{rag90b} Raga, A. C., Cant\'o, J., Binette, L., \& Calvet, N.
  1990b, ApJ, 364, 601

\bibitem[{Raga et al.} (1997)]{rag97} Raga, A. C., Cant\'o, J., Curiel, S., Noriega-Crespo, A.,
  \& Raymond, J. C. 1997, RMxAA, 33, 157

\bibitem[{Raga \& Cant\'o} (2003)]{rag03} Raga, A. C., \& Cant\'o, J. 1994,
  A\&A, 412, 745

\bibitem[{Raga et al.} (2010)]{rat10} Raga, A. C., Cant\'o, J., Esquivel, A., Rodr\'\i guez-Gonz\'alez, A.,
Vel\'azquez, P. F. 2010, in {\it Highlights in Astronomy}, Vol. 15, p. 256

\bibitem[{Raga et al.} (2011)]{rag11} Raga, A. C., Reipurth, B., Cant\'o, J., Sierra-Flores, M. M.,
  \& Guzm\'an, M. V. 2011, RMxAA, 47, 425

\bibitem[{Raga et al.} (2012)]{rag12} Raga, A. C., Noriega-Crespo, A., Rodr\'\i guez-Gonz\'alez, A.,
  Lora, V., Stapelfeldt, K. R., \& Carey, S. J. 2012, ApJ, 748, 103

\bibitem[{Raga et al.} (2015a)]{rag15a} Raga, A. C., Reipurth, B., Castellanos-Ram\'\i rez, A.,
  Chiang, H.-F., \& Bally, J. 2015a, ApJ, 798, L1

\bibitem[{Raga et al.} (2015b)]{rag15b} Raga, A. C., Reipurth, B., Castellanos-Ram\'\i rez, A.,
  Chiang, H.-F., \& Bally, J. 2015b, AJ, in press


%\bibitem[Raga\& Binette 1991]{rag91} Raga, A. C., Binette, L. 1991, RMxAA, 22, 265

%\bibitem[Raymond 1979]{ray79} Raymond, J. C. 1979, ApJS, 39, 1

%\bibitem[Reipurth et al. 2002]{rei02} Reipurth, B., Heathcote, S., Morse, J., Hartigan, P.,
%\& Bally, J. 2002, AJ, 123, 362

%\bibitem[Reipurth et al. 1997]{rei97} Reipurth, B., Hartigan, P., Heathcote, S., Morse, J.,
%\& Bally, J. 1997, AJ, 114, 757

%\bibitem[Rodr\'\i guez et al. 2000]{rod00} Rodr\'\i guez, L. F., et al. 2000, AJ, 119, 882

%\bibitem[Schwartz et al. 1994]{sch93} Schwartz, R. D., et al. 1993, AJ, 106, 740

%\bibitem[Solf et al. 1988]{sol88} Solf, J., B\"ohm, K. H., \& Raga, A. C.
%1988, ApJ, 334, 229

%\bibitem[Solf et al. 1991]{sol91} Solf, J., Raga, A. C., B\"ohm, K. H.,
%Noriega-Crespo, A. 1991, AJ, 102, 1147

\bibitem[{Szyszka et al.} (2011)]{szy11} Szyszka, C., Zijlstra, A. A., \& Walsh, J. R. 2011,
  MNRAS, 416, 715

\bibitem[{Strom et al.} (1985)]{str85} Strom, S. E., Strom, K. M., Grasdalen, G. L.,
Sellgren, K., Wolff, S., Morgan, J., Stocke, J., \& Mundt, R. 1985, AJ, 90, 2281


\end{thebibliography}
\end{document}